\documentclass[conference]{IEEEtran}
\IEEEoverridecommandlockouts

\usepackage[T1]{fontenc}
\usepackage[utf8]{inputenc}
\usepackage{lmodern}
\usepackage{doi}

\usepackage{cite}
\usepackage{listings}
\usepackage{xcolor}
\usepackage{amsmath,amssymb,amsfonts}
\usepackage{algorithmic}
\usepackage{graphicx}
\usepackage{caption}
\usepackage[skip=5pt]{caption}
\usepackage{mdframed}
\usepackage{textcomp}
\usepackage{xcolor}
\usepackage{minted}

\def\BibTeX{{\rm B\kern-.05em{\sc i\kern-.025em b}\kern-.08em
    T\kern-.1667em\lower.7ex\hbox{E}\kern-.125emX}}
\begin{document}

\title{Automated Program Repair Based on REST API Specifications Using Large Language Models
}

\author{
\IEEEauthorblockN{Katsuki Yamagishi}
\IEEEauthorblockA{\textit{Ritsumeikan University}\\
Ibaraki, Japan \\
is0585fr@ed.ritsumei.ac.jp}
\and
\IEEEauthorblockN{Norihiro Yoshida}
\IEEEauthorblockA{\textit{Ritsumeikan University}\\
Ibaraki, Japan \\
norihiro@fc.ritsumei.ac.jp}
\and
\IEEEauthorblockN{Erina Makihara}
\IEEEauthorblockA{\textit{Ritsumeikan University}\\
Ibaraki, Japan \\
makihara@fc.ritsumei.ac.jp}
\and
\IEEEauthorblockN{Katsuro Inoue}
\IEEEauthorblockA{\textit{Ritsumeikan University}\\
Ibaraki, Japan \\
inoue-k@fc.ritsumei.ac.jp}
}

\maketitle

\begin{abstract}
Many cloud services provide REST API accessible to client applications. However, developers often identify specification violations only during testing, as error messages typically lack the detail necessary for effective diagnosis. Consequently, debugging requires trial and error.
This study proposes dcFix, a method for detecting and automatically repairing REST API misuses in client programs. In particular, dcFix identifies non-conforming code fragments, integrates them with the relevant API specifications into prompts, and leverages a Large Language Model (LLM) to produce the corrected code.
Our evaluation demonstrates that dcFix accurately detects misuse and outperforms the baseline approach, in which prompts to the LLM omit any indication of code fragments non conforming to REST API specifications.
\end{abstract}


\section{Introduction}
Recently, cloud services, such as Infrastructure as a Service (IaaS), Platform as a Service (PaaS), and Software as a Service (SaaS), have rapidly and widely been adopted.
Various cloud services provide a REpresentational State Transfer (REST) API \cite{REST}, an architectural style that enables web-based access for third-party applications, not limited to those developed by the service providers themselves.
To access and manipulate resources managed by cloud services, clients send HTTP requests to the REST API.
These requests typically specify the resource path as the endpoint and the desired action as the HTTP method (e.g., GET, POST).
In most cases, clients convey metadata (e.g., data format) in HTTP headers and transmit payloads (e.g., text data) in the HTTP body.

In REST API client development, it is common for developers to realize that their program does not comply with the API specifications only during testing, upon examining API responses.
The error code and messages included in the responses do not often provide sufficient information to identify that the API specification has been violated.
Consequently, developers frequently debug by iteratively sending requests and analyzing the responses.

Previous research has addressed the API misuse detection \cite{Li2024, Ren2000, Amann2019} and automated repair \cite{Kechagia2022, Nielebock2017, Zhang2022} of Java API.
These studies have managed to detect Java API calls in the source code that violate the API specifications.
However, the Java API differs significantly from the REST API in its invocation model on the client side.
In Java, compliance with specifications is required at the method-call level.
In contrast, REST API clients must ensure that the specified endpoints, request headers, and request bodies are consistent with the API specifications.
Consequently, it is difficult to apply these Java-based misuse detection techniques to the REST API directly.

This study proposes {\sf dcFix} to detect and automatically repair REST API misuses in client programs.
{\sf dcFix} first identifies code fragments (hereafter ``deviation points'') that do not comply with the API specifications.
Subsequently, it generates a prompt that includes the deviation point and the unsatisfied part of the specification (hereafter ``unsatisfied specification''), which is then provided to an LLM (Large Language Model).
Finally, the LLM is used to automatically repair the misuse based on the generated prompt.

In our evaluation, we applied {\sf dcFix} to misuse cases collected from the {\sf SwitchBot} and {\sf Philips Hue} APIs.
First, we investigated the ability of {\sf dcFix} to detect deviation points and found that it identified them successfully in most cases.
Subsequently, we compared {\sf dcFix} to a baseline approach in which prompts to the LLM omit unsatisfied specifications or deviation points.
The results show that {\sf dcFix} corrected more misuse cases than the baseline approach.

The main contributions of this study are as follows:
\begin{itemize}
\item We propose {\sf dcFix}, a hybrid approach that, rather than simply using an LLM, analyzes the program against the REST API specification to identify unsatisfied specifications and deviation points, and includes them in the LLM prompt.

\item Compared to the baseline approach, which omits unsatisfied specifications or deviation points in LLM prompts, {\sf dcFix} can automatically repair more REST API misuses.

\item For the evaluation of {\sf dcFix}, we collected REST API misuse examples from OSS and published them on the website\footnote{https://zenodo.org/records/16556822} alongside the {\sf dcFix} implementation.

\end{itemize}

\section{Motivating Example}
Listing 1 presents an example of a code fragment containing the SwitchBot API misuse, which occurs in the request header section from lines 6 to 10. In addition to the {\sf Authorization} field, the request header should include fields, such as {\sf sign, t, nonce}. 


{\small
\begin{lstlisting}[caption=Misuse Example of the SwitchBot API, label=LLM, language=python]
             :
def get_device_list() -> json:
    url = f'{API_URL}/devices'
    response = requests.get(url, headers=HEADERS)
    return response.json()
    HEADERS = {
    'Authorization': OPEN_TOKEN,
    'Content-Type': 'application/json; charset=utf-8'
    #The request header is missing the required values for sign, t, and nonce.
    }
            :
\end{lstlisting}
}

Such misuses are difficult to detect without carefully comparing the implementation against the official REST API specifications.
In practice, developers often have to rely on trial and error: executing the program, interpreting short error responses from the server, identifying potential issues in endpoints or request headers, and repeatedly rerunning the program to confirm fixes.
This process is time-consuming and error-prone, especially when the specifications are complex or have recently changed.
Automated support that can highlight missing or incorrect fields—such as {\sf sign}, {\sf t}, and {\sf nonce}—based on up-to-date API specifications would greatly reduce the developer's burden and improve the reliability of client-side code.






\section{{dcFix}}\label{sec3}
\begin{figure*}[t]
\setlength{\abovecaptionskip}{8pt}
\vspace{10pt}
    \centering
    \vspace{-3.6\baselineskip}
   \includegraphics[width=\textwidth]{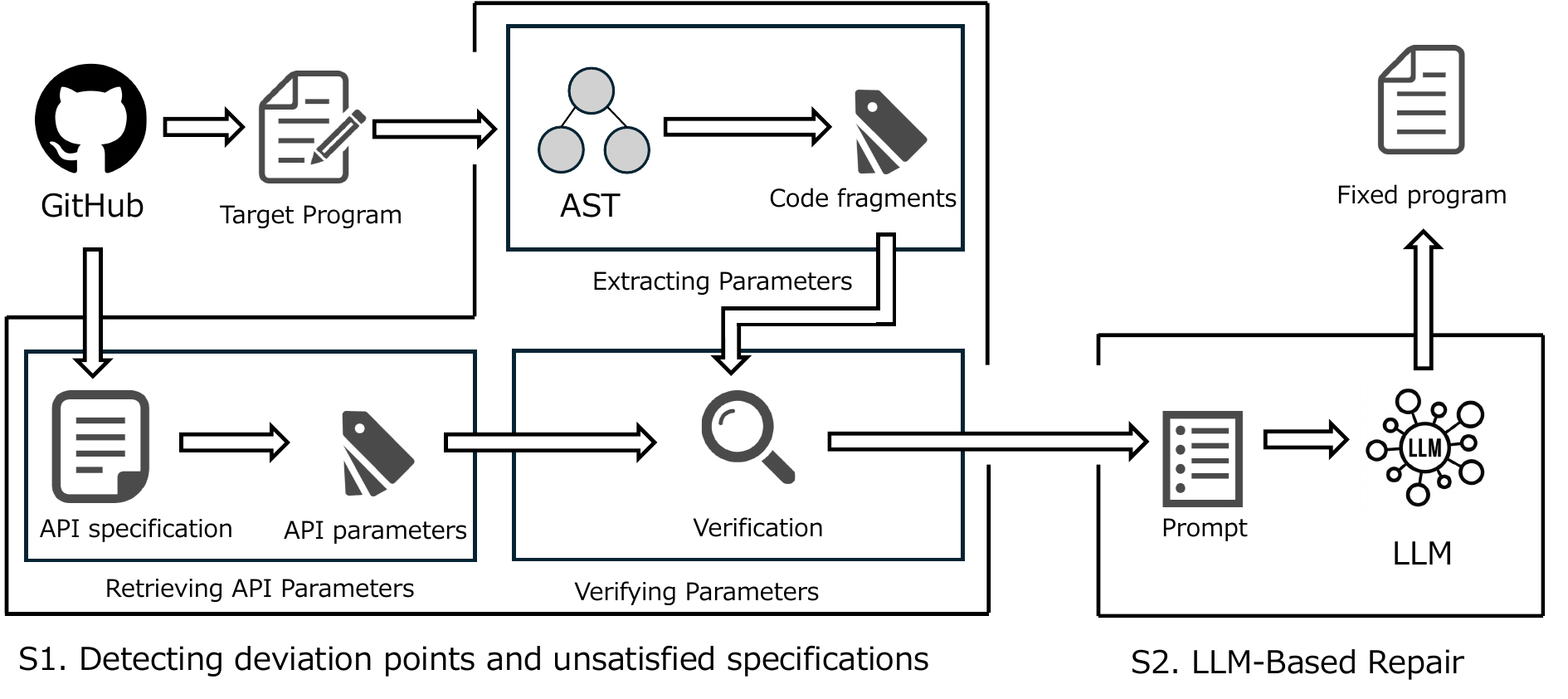}
    \caption{An Overview of {\sf dcFix}}
    \label{fig3}
\end{figure*}

In this section, we propose the {\sf dcFix} method that leverages an LLM to correct programs that deviate from the REST API specifications.
Figure \ref{fig3} illustrates an overview of {\sf dcFix}.
\begin{figure*}[t]
\setlength{\abovecaptionskip}{8pt}
\vspace{30pt}
    \centering
    \vspace{-3.6\baselineskip}
   \includegraphics[width=0.75\textwidth]{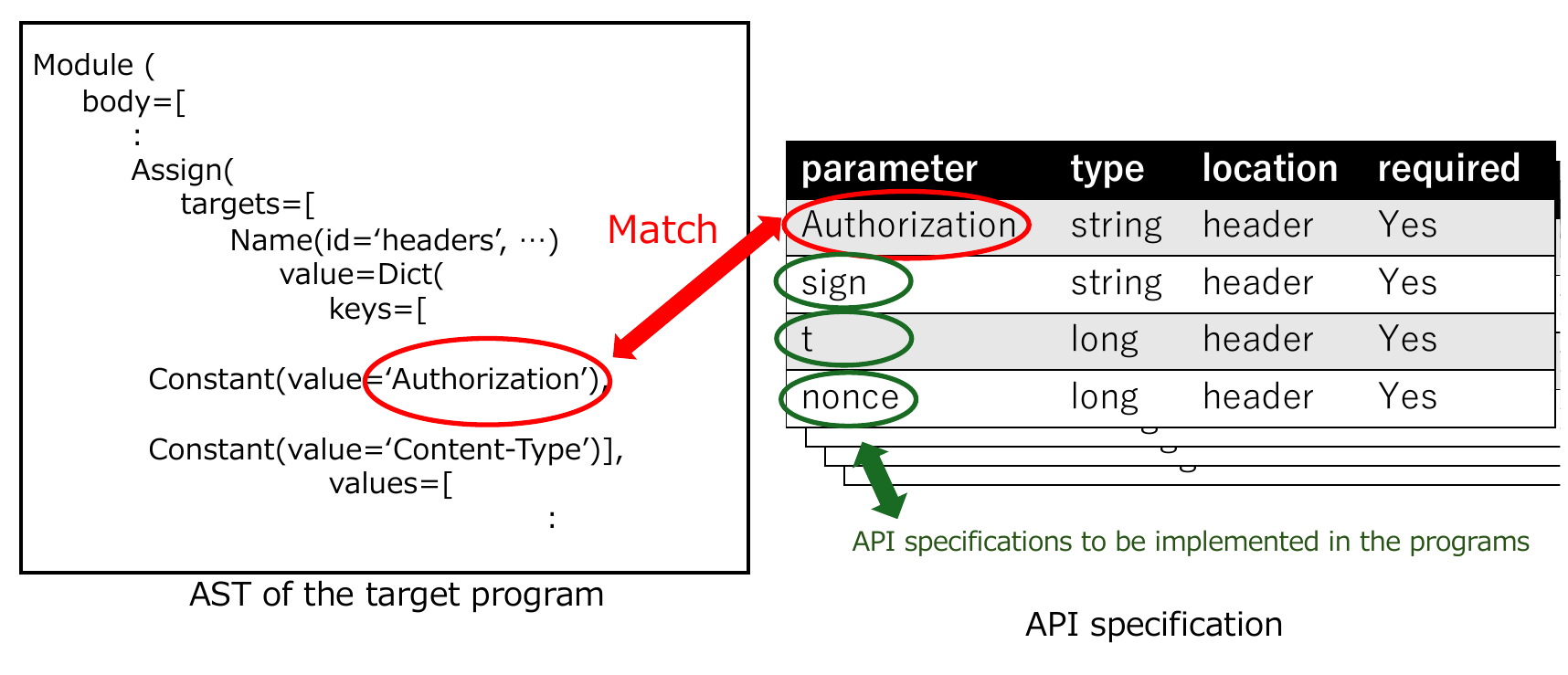}
    \caption{Matching between a deviation point (left) and its corresponding unsatisfied specification (right)}
    \label{fig4}
\end{figure*}
It consists of the following steps:
\begin{enumerate}
\item[S1] {\bf Detecting deviation points and unsatisfied specifications}
\item[S2] {\bf LLM-Based Repair}
\end{enumerate}
In the following subsections, we explain each step.

\subsection{Detecting Deviation points and unsatisfied specifications}

First, {\sf dcFix} identifies code fragments invoking the target REST API by extracting the API’s domain name from its specification (YAML or JSON format) and searching the code for string literals containing that domain. Next, it performs data flow analysis on those literals to identify fragments that define endpoints and declare request headers/bodies.

Then, {\sf dcFix} performs static analysis on the target program to verify that the declared endpoint is included in the REST API specification and that attributes required to invoke the endpoint are declared in the request headers/bodies. For request bodies, if the endpoint’s specification defines particular attribute values, {\sf dcFix} checks that those values are actually assigned.

Figure \ref{fig4} shows an example of detecting unsatisfied specifications. The left side illustrates the target program’s AST, and the table on the right lists the attributes required by the API calls in the program. The table shows that four attributes (i.e., {\sf Authorization}, {\sf sign}, {\sf t}, {\sf nonce}) are required, but the AST shows that only Authorization is declared. In this case, {\sf dcFix} detects {\sf sign}, {\sf t}, and {\sf nonce} as unsatisfied specifications. Furthermore, it identifies the code corresponding to the AST on the left as a deviation point.


The deviation points to be specified vary by parameter type.
Below, we describe the deviation points for each parameter type in detail:
\begin{enumerate}
       \item \textbf{Endpoint:} Whether the URL used to invoke the API matches the URL specified in the API specification.
       \item \textbf{Request Headers:} Whether each required attribute in the API request headers, as specified in the API specification, is defined in the program. The actual values of these attributes are not considered.
       \item \textbf{Request Body:} Whether all required attributes defined in the API specification are present in the program. If the API specification provides specific values for certain attributes, those values are used for comparison; otherwise, the values are not considered.
\end{enumerate}

When {\sf dcFix} detects a misuse, such as an endpoint excluded by the REST API specification or missing required attributes, it marks the corresponding code fragment as a deviation point and identifies the related unsatisfied specification causing the deviation.


Based on the detected deviation points and unsatisfied specifications, {\sf dcFix} constructs a LLM prompt using a predefined template designed to guide the correction.
This prompt is then submitted to the LLM, which automatically generates a corrected version of the deviation point.

\subsection{LLM-Based Repair}
Correction was performed by inputting a prompt into the LLM that included the location information and attribute values of the detected deviation points. This prompt is generated based on a predefined template as shown in Figure \ref{prompt} and incorporates the deviation point information obtained in Step S1.
The prompt includes a variable corresponding to the deviation point (i.e., the attribute defined in the specification) along with the proposed correction, which is provided as an input to the LLM. The template is designed to be concise and focused on the deviation and its correction, minimizing the influence of unrelated code on the LLM's output.

Listing~\ref{prompt} illustrates the prompt template used for repairing misuse cases.
The template includes a placeholder for a program that contains a deviation point, typically starting at Line 10, and explicitly incorporates insufficient specification details intended to guide the model in generating a fix that complies with the API specification.

\begin{lstlisting}[caption=A prompt template used by {\sf dcFix} to generate repair instructions, label=prompt, breakindent=0pt]
"""
You are an AI specialized in fixing programming bugs. 
Please modify the code provided by the user according to the given instructions. 
When making changes, avoid altering the overall structure of the program. 
Focus only on simple fixes that address the bug and its related parts. 
Only return the corrected code inside a single code block. 
Do not include any explanations or additional text.
Please Update the code to match the latest API specifications the following code.
"""
${target program}
\end{lstlisting}

\section{Case Study}\label{sec4}
This section outlines the preparation of misuse cases for the experiment and presents the corresponding experimental results. For the LLM-based experiments, we prepared programs with code fragments containing deviations and generated corresponding prompts. The LLM performed the correction task five times per prompt; a correction was considered successful if at least one attempt produced a correct result.

In this experiment, we defined the following two research questions:
\begin{description}
    \item [RQ1:] {\bf Does {\sf dcFix} detect deviation points from the specification?}\\This RQ investigates how many misuse cases among the collected API examples were successfully detected by the {\sf dcFix}.
    \item [RQ2:] {\bf Does {\sf dcFix} improve the fix rate of misuse repair over the baseline approach?}    
\\This RQ examines whether {\sf dcFix} achieves a higher fix rate in repairing misuse cases compared to the baseline approach, which does not include unsatisfactory specifications or deviation points in the prompt.

\end{description}
\begin{table}[t]
    \centering
    \caption{Statistics of the target misuses}
    \begin{tabular}{c|cc}
                    \hline
                    \hline
                    &Philips Hue API&SwitchBot API
                    \\\hline
                    Endpoint & 14 & 5\\
                    Request Headers & 0 & 4\\
                    Request Body & 7 & 1\\\hline
                    Total & 21 & 10
                    \\\hline
    \end{tabular}
    \label{table1}
\end{table}
\begin{table}[t]
    \centering
    \caption{Detection results of deviation points using {\sf dcFix}}
    \begin{tabular}{c|cc}
                    \hline
                    \hline
                    &Philips Hue API&SwitchBot API
                    \\\hline
                    Endpoint & 12/14 & 5/5\\
                    Request Headers & n/a & 2/4\\
                    Request Body & 7/7 & 1/1\\\hline
                    Total & 19/21 & 8/10
                    \\\hline
    \end{tabular}
    \label{table2}

\end{table}
\begin{table}[t]
    \caption{Comparison of fix rates between {\sf dcFix} and the baseline approach}
    \centering
    \begin{tabular}{c|cc|cc}
        \hline\hline
         & \multicolumn{2}{c|}{Philips Hue API} & \multicolumn{2}{c}{SwitchBot API} \\
        \cline{2-5}
         & {\sf dcFix} & baseline & {\sf dcFix} & baseline \\
        \hline
        Endpoint         & 8/12 & 1/12 & 4/5 & 0/5 \\
        Request Headers  & n/a  & n/a & 1/2 & 0/2 \\
        Request Body     & 5/7  & 0/7 & 1/1 & 1/1 \\\hline
        Total            & 13/19 & 1/19 & 6/8 & 1/8 \\
        \hline
    \end{tabular}
    \label{table3}
\end{table}

\subsection{Dataset}
The experiments used the dataset listed in Table \ref{table1}, which was collected from GitHub. 
Table \ref{table1} summarizes, for each API, how many misuse cases were associated with each category of unsatisfactory specification.

The repository targeted for the SwitchBot API is {\sf https://github.com/OpenWonderLabs/SwitchBotAPI}, and the data were collected from issues that had already been closed. To establish selection criteria, we collected data from programs that violated API specifications. For the SwitchBot API, no correction examples were found in issue reports; thus, we gathered cases where API descriptions or specification parameters had been modified in commits. We manually extracted programs containing code fragments that were inconsistent with the specifications.

For the Philips Hue API, we targeted commits from repositories that included the newly introduced endpoint parameter {\sf /clip/v2} that was added during the version update, focusing specifically on commits in which {\sf /clip/v2} was modified in the update.

In this study, we used these two sources for our dataset. All tasks related to the preparation of the dataset were performed by the first author.
\subsection{Results}
This section discusses the results of the RQs used in the experiments. 
For RQ1, as shown in Table \ref{table2}, deviations from the SwitchBot API specifications were identified in all 5 endpoint cases, 2 out of 4 request header cases, and 1 request body case.
In contrast, for the Philips Hue API, deviations were detected in 12 out of 14 endpoint cases and all 7 requested body cases.
\begin{mdframed}
\textbf{Answer to RQ1:} 
{\sf dcFix} was able to detect a large number of deviations points in the dataset.
It detected 19 out of 21 deviations for the Philips Hue API and 8 out of 10 deviations for the SwitchBot API. 
\end{mdframed}

Table \ref{table3} presents the number of successfully repaired cases for both the Philips Hue API and the SwitchBot API under two conditions: using the baseline approach, which prompts the LLM without including unsatisfactory specifications or deviation points, and using {\sf dcFix} in combination with the LLM.
For the Philips Hue API, incorporating {\sf dcFix} led to an increase of 12 successful repairs, while for the SwitchBot API, the use of dcFix resulted in 5 additional successful repairs.


As shown in Listing \ref{LLM}, this modification was generated by {\sf dcFix}, in response to updated requirements in the SwitchBot API.
Specifically, the API version was updated from 1.0 to 1.1, and missing request headers (i.e., {\sf nonce}, {\sf t}, and {\sf sign}) were added accordingly.

\begin{mdframed}
\textbf{Answer to RQ2:} {\sf dcFix} achieved higher fix rates compared to the baseline approach for both the Philips Hue and SwitchBot APIs.
\end{mdframed}
{\small
\begin{lstlisting}[caption=Repair example of a SwitchBot API misuse using {\sf dcFix}, label=LLM, language=python]
                :
  import requests
  import json
  import os
  from dotenv import load_dotenv
  load_dotenv()

  OPEN_TOKEN = os.getenv('OPEN_TOKEN')

  API_HOST = 'https://api.switch-bot.com'

- DEBIVELIST_URL = f"{API_HOST}/v1.0/devices"
+ DEBIVELIST_URL = f"{API_HOST}/v1.1/devices" 
# <-- Update to v1.1

# Request information
  HEADERS = {
    'Authorization': OPEN_TOKEN,
    'Content-Type': 'application/json; charset=utf8'
+   'nonce': 'nonceValue',  # <-- add nonce value
+   't': 'tValue',  # <-- add t value
+   'sign': 'signValue'  # <-- add sign value
}
                :
\end{lstlisting}
}
\section{Related Work}
Prior studies have investigated the detection \cite{Li2024, Ren2000, Amann2019} and automatic repair \cite{Kechagia2022} of API misuses, particularly for Java API. Sven et al. proposed MUDetect, a tool that detects and ranks API misuse using an API Usage Graph, a graph-based representation of API usage patterns \cite{Amann2019}. Ren et al. introduced a method that constructs graphs representing API usage constraints and detects misuse instances based on these graphs \cite{Ren2000}. Li et al. extracted API usage constraints from clients, libraries, and Java API documentation to construct API usage constraint graphs for misuse detection \cite{Li2024}. Kechagia et al. evaluated the effectiveness of existing automatic repair tools in addressing Java API misuse instances \cite{Kechagia2022}.However, Java API and REST API differ significantly in their client invocation mechanisms. For Java API, the correctness of method invocations is crucial, whereas for REST API, the correctness of elements, such as endpoints, request headers, and request bodies written in the client code, must conform to the specification. Consequently, applying Java API misuse detection techniques directly to REST API is challenging.
In the domain of REST API, several studies have focused on verifying the REST API provided by cloud services \cite{RESTler, Huang2024}. 

Atlidakis et al. developed a fuzzing tool named RESTler, which automatically generates API call sequences by analyzing dependencies among REST API calls. This tool has been applied to verify REST API offered by services, such as GitLab, Microsoft Azure, and Office 365 \cite{RESTler}. Huang et al.  proposed a technique using program analysis to verify whether REST API provided by cloud services conform to their specifications \cite{Huang2024}. In contrast, the present study proposes an automatic repair method targeting client-side code that invokes REST API.

Xia et al. demonstrated that LLM outperform traditional automatic bug repair tools in generating code fixes \cite{Chunqiu}. Jin et al. further showed that combining static analysis tools with LLM yields more effective automatic repair than using LLM alone \cite{Jin2023}. Although our approach similarly integrates program analysis with LLM, the types of unmet specifications and deviations targeted here are difficult to detect using conventional static analysis tools.

\section{Summary}\label{sec7}
This study aims to automatically correct client-side programs that violate REST API specifications using LLMs.
In the {\sf dcFix} method, unsatisfied specifications are extracted from the official API documentation, while corresponding program elements are retrieved from the client code.
By comparing the two, {\sf dcFix} identifies deviation points—code fragments that do not satisfy the required specifications.
A prompt is then generated by incorporating the location and content of these deviation points, along with relevant surrounding code.
This prompt is provided as input to the LLM to perform automatic bug correction.

In the applicability experiment, we applied this method to the SwitchBot API and Philips Hue API to evaluate its effectiveness in repairing misuse cases.
The results showed that prompts containing information about deviation points and unsatisfied specifications led to a higher fix rate compared to prompts that lacked such information.

In our case study, there were cases where the LLM failed to generate correct fixes because it had not been trained on the latest REST API specifications.
To address this limitation, a promising future direction is to incorporate Retrieval-Augmented Generation (RAG) techniques to explicitly provide the LLM with up-to-date specification details.
By doing so, the repair accuracy and efficiency could be further improved, especially in scenarios where API specifications frequently change.


\bibliographystyle{IEEETran}
\bibliography{reference}
\end{document}